\documentclass[prl,aps,twocolumn]{revtex4}

\usepackage{amsmath}
\usepackage{epsfig}
\usepackage{amsfonts}

\begin{document}

\title{Incommensurate magnetic order in Sm$_3$BWO$_9$ with the distorted kagome lattice}

\author{K. Y. Zeng$^{1,2}$}
\author{F. Y. Song$^{3}$}
\author{L. S. Ling$^{1}$}
\author{W. Tong$^{1}$}
\author{Shiliang Li$^{4,5}$}
\author{Z. M. Tian$^{3}$}
\email{tianzhaoming@hust.edu.cn}
\author{Long Ma$^{1}$}
\email{malong@hmfl.ac.cn}
\author{Li Pi$^{1,2}$}

\affiliation{$^{1}$Anhui Province Key Laboratory of Condensed Matter Physics at Extreme Conditions, High Magnetic Field Laboratory, Chinese Academy of Sciences, Hefei 230031, China\\
$^{2}$ Hefei National Research Center for Physical Sciences at the Microscale, University of Science and Technology of China, Hefei 230026, China\\
$^{3}$ School of Physics and Wuhan National High Magnetic Field Center, Huazhong University of Science and Technology, Wuhan 430074, PR China\\
$^{4}$ Beijing National Laboratory for Condensed Matter Physics, Institute of Physics, Chinese Academy of Sciences, Beijing 100190, China\\
$^{5}$ School of Physical Sciences, University of Chinese Academy of Sciences, Beijing 100190, China}

\date{\today}

\begin{abstract}
  We investigate the magnetic ground state of Sm$_3$BWO$_9$ with the distorted kagome lattice. A magnetic phase transition is identified at $T_N=0.75$ K from the temperature dependence of specific heat. From $^{11}$B nuclear magnetic resonance (NMR) measurements, an incommensurate magnetic order is shown by the double-horn type spectra under a $c$-axis magnetic field. While, absence of line splitting is observed for field oriented within the $ab$-plane, indicating the incommensurate modulation of the internal field strictly along $c$-axis. From the spin dynamics, the critical slowing down behavior is observed in the temperature dependence of $1/T_1$ with $\mu_0H\perp c$-axis, which is completely absent in that with $\mu_0H||c$-axis. Based on the local symmetry of $^{11}$B sites, we analyze the hyperfine coupling tensors and  propose two constraints on the possible magnetic structure. The single ion anisotropy should play an important role in the determination of the contrasting ground states of Sm$_3$BWO$_9$ and Pr$_3$BWO$_9$.
\end{abstract}

\maketitle

Frustrated magnetism gains its unfailing research interest in the condensed matter community for the existence of exotic states or spin excitations\cite{Lacroix_book, Balents_Nature_2010}. The antiferromagnetic Heisenberg spins located on corner-sharing triangles (kagome lattice) form the archetype frustrated magnet, whose intrinsic ground state and novel spin excitation is still elusive for decades. The infinite degeneracy resulting from strong frustration effect leads to various exotic states, such as quantum spin liquids\cite{Balents_Nature_2010,Zhou_RMP_2017}, spin-orbital liquids\cite{Schaffer_PRB_88_174405}, kagome spin ice\cite{Dun_PRL_116_157201,Wills_PRB_66_144407}, spin nematicity\cite{Picot_PRB_91_064415}, et al. For the quantum ($S=1/2$) case, considerable divergence exists among the researchers about its true ground state, which is widely believed to be magnetically disordered\cite{Singh_PRB_76_180407,Jiang_PRL_101_117203,Nakano_JPSJ_80_053704}. In the classical limit ($S=\infty$), the non-linear ordered state with 120$^o$ between neighbor spins are selected by thermal fluctuations out of the infinitely degenerate spin configurations\cite{Chalker_PRL_68_855,Reimers_PRB_48_9539,Robert_PRL_101_117207}. This release of degeneracy due to thermal fluctuations is known as "order by disorder"\cite{Chalker_PRL_68_855}. The spin dynamics are mainly contributed by the zero-energy soft mode which can be viewed as two spin sublattices rotating about the main-axis defined by the third spin\cite{Reimers_PRB_48_9539}.

The spin state of magnetic ions has profound influence on the ground state of frustrated magnets. The $4f$-rare earth ions share similar chemical properties but possess diverse spin-orbit entangled effective moments. Thus, the rare earth based frustrated magnets supply valuable material platform for realizing various exotic quantum or classical magnetic ground states and excitations. One typical example is the RE$_3$Sb$_3$M$_2$O$_{14}$(RE=Pr, Nd, Sm-Yb, M=Zn, Mg) compounds with kagome lattice\cite{Dun_PRL_116_157201}, whose ground states are identified to be kagome spin ice, dipolar spin order, et al. Interestingly, possible gapless spin liquid state is proposed for Tm$_3$Sb$_3$Zn$_2$O$_{14}$ by the $\mu$SR study\cite{Ding_PRB_98_174404}. However, the random mixing of Tm$^{3+}$ and Zn$^{2+}$ is proved to exist\cite{Ma_PRB_102_224415}, which leads to the spin-liquid-like behavior.

Recently, a new family of RE$_3$BWO$_9$ (RE=Pr, Nd, Gd-Ho) are synthesized by some of us\cite{Ashtar_IC_59_5368}. The RE$^{3+}$ ions lie on the distorted kagome lattice in $ab$-plane, which stacks along $c$-axis in an $AB$-type fashion. The lattice distortion is realized by alternatively arranging two kinds of regular triangles with different side length around the center hexagon. Inherent site-mixing disorder is avoided in these boratotungstates thanks to the large difference of ionic radii and coordination number, which is very critical in magnetic frustration studies. Dominant antiferromagnetic couplings between RE$^{3+}$ ions are revealed by magnetic susceptibility measurements, while no magnetic order is observed down to $T=2$ K\cite{Ashtar_IC_59_5368}. In Pr$_3$BWO$_9$, the nuclear magnetic resonance (NMR) study reveals a novel cooperative ground state with persistent collective spin excitations down to $T=0.09$ K, far below its Curie-Weiss temperature\cite{Zeng_PRB_104_155150}. However, no further studies on the ground state of other members of RE$_3$BWO$_9$ are reported.

In this paper, we study the magnetic ground state of Sm$_3$BWO$_9$. A magnetic phase transition is identified at $T_N=0.75$ K from the temperature dependence of specific heat under zero magnetic field. From $^{11}$B NMR results, the low temperature magnetic ordering leads to the double-horn type spectra with $\mu_0H||c$-axis, characteristic for the single-$\overrightarrow{q}$ modulation of the incommensurate order. With the field applied within the $ab$-plane, line splitting due to the long range order is absent, indicating the internal field strictly along $c$-axis. For the spin dynamics, the critical slowing down behavior around $T_N$ shows up from $1/T_1(T)$ under the in-plane magnetic field, while is not observed in $1/T_1(T)$ with the field applied along $c$-axis. From analysis of the nuclear hyperfine coupling tensor based on the local crystalline symmetry, we propose two constraints on the possible magnetic structure. With the magnetic field increasing, the magnetic ordering is gradually suppressed, and becomes undetectable for the studied temperature range above $\mu_0H=5$ T.

Single crystals of Sm$_3$BWO$_9$ with high quality were grown by the conventional flux method\cite{Ashtar_IC_59_5368}. For the NMR study, we chose single crystals with typical dimensions of $1\times1\times1.5$ mm$^3$. The NMR measurements are conducted on $^{11}$B nuclei ($\gamma_n=13.655$ MHz/T, $I=3/2$) with a phase-coherent spectrometer. The spectra are obtained by summing up or integrating the spectral weight by sweeping the frequency or magnetic field with the other fixed. The spin-lattice relaxation rate ($1/T_1$) is measured by the inversion-recovery sequence, and obtained by fitting the time dependence of nuclear magnetization of the central transition or satellites.

\begin{figure}
\includegraphics[width=8cm, height=8cm]{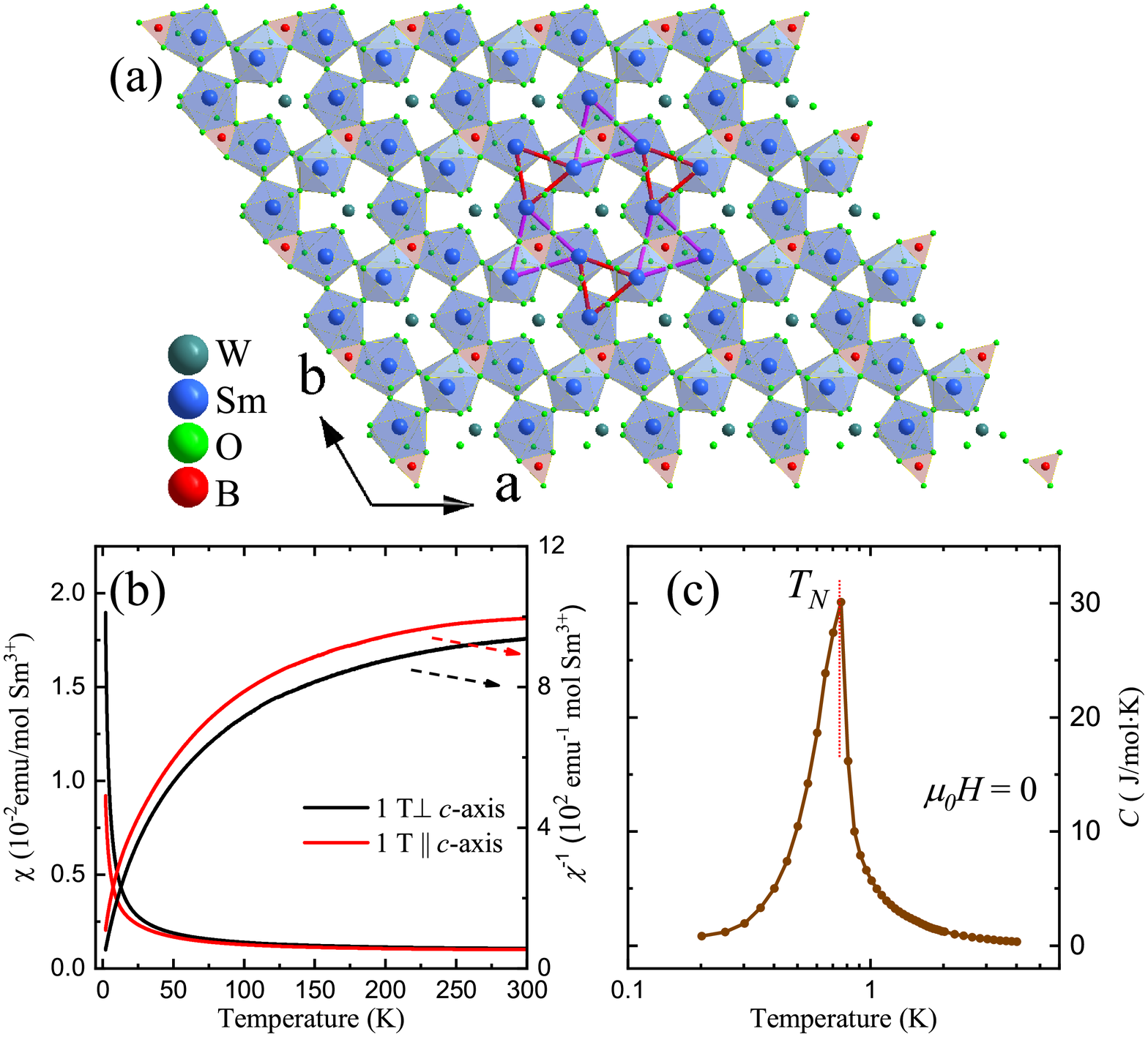}
\caption{\label{char1}(color online) (a) The crystalline structure of Sm$_3$BWO$_9$ as seen against $c$-axis. The regular triangles with different side length from the distorted kagome lattice are marked with red and pink lines respectively. (b) The temperature dependence of $dc$-susceptibility and its reciprocal under different field directions. (c) The specific heat under zero field as a function of temperature.
}
\end{figure}

The Sm$_3$BWO$_9$ crystalizes in the hexagonal structure with the P6$_3$ space group. In Fig.\ref{char1}(a), we show the crystalline structure of Sm$_3$BWO$_9$ as seen against $c$-axis. The magnetic SmO$_8$ polyhedra are interconnected by the nonmagnetic WO$_6$ and BO$_3$ polyhedra in the corner- or edge-sharing manner\cite{Ashtar_IC_59_5368}. Regular Sm$^{3+}$-triangles with different edge length are marked respectively with red and pink bonds. For Sm$_3$BWO$_9$, there are only one Wyckoff position respectively for Sm, W, and B sites, and three O positions. The direct edge-sharing of BO$_3$ triangles and SmO$_8$ polyhedra leads to strong hyperfine coupling between $^{11}$B nuclei and magnetic Sm$^{3+}$ ions.

The temperature dependence of $dc$-susceptibility and the reciprocal of $dc$-susceptibility are shown in Fig.\ref{char1}(b). The $dc$-susceptibility is measured with a commercial SQUID with the magnetic field applied parallel or perpendicular to the $c$-axis. For both field directions, the susceptibility show a Curie-Weiss-like upturn behavior upon cooling, typical for localized spin systems. The temperature dependence of heat capacity is shown in Fig.\ref{char1}(c), without substraction of phonon or Schottky anomaly contributions. The obvious $\lambda$-peak centered at $T_N=0.75$ K clearly indicates the phase transition to the magnetically ordered state. To gain further information about the ground state, we perform detailed NMR study on the low temperature magnetic property of Sm$_3$BWO$_9$.

\begin{figure}
\includegraphics[width=8cm, height=7.5cm]{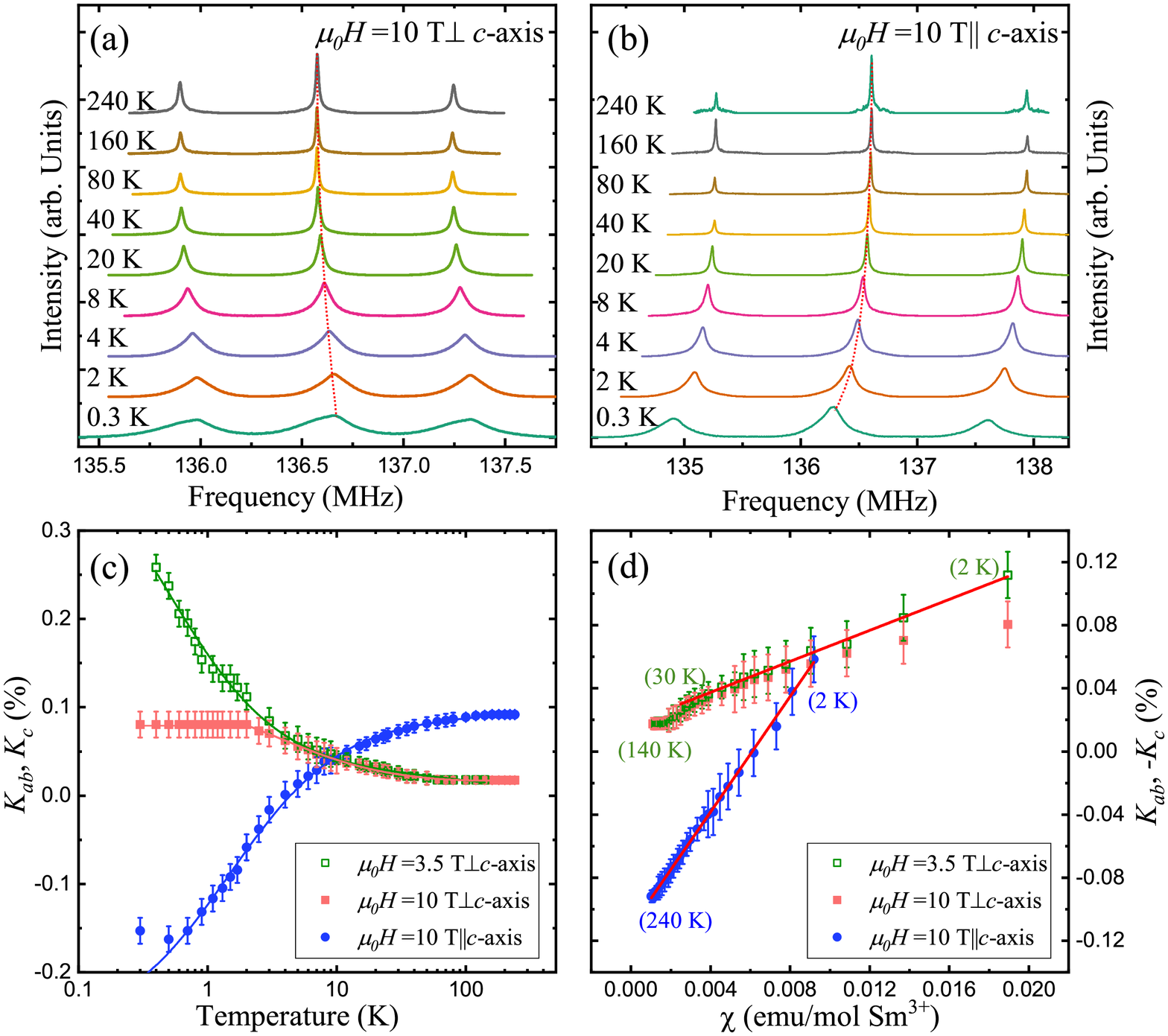}
\caption{\label{knight2}(color online)
The frequency-swept $^{11}$B NMR spectra at different temperatures with 10 T field applied perpendicular to(a) or along $c$-axis(b). The spectral intensity is Boltzmann-corrected to get rid of the influence of temperature.
(c): The temperature dependence of $^{11}$B Knight shift $^{11}K$. (d) The Knight shift plotted against $dc$-susceptibility with the temperature as an explicit parameter.
}
\end{figure}

In Fig.\ref{knight2}(a) and (b), we show typical $^{11}$B NMR spectra at different temperatures with 10 T field applied parallel or perpendicular to the $c$-axis. The spectra are composed by three peaks, one central transition and double satellites located symmetrically at both sides. For $^{11}$B with nuclear spin $I=3/2>1/2$ lying on the lattice with symmetry lower than cubic, the non-zero nuclear quadruple moment $Q$ interacts with the electric field gradient (EFG) of local surrounding charges. The first order correction to the resonance frequency determined by Zeeman splitting can be written as, $\nu_m^{(1)}=\nu_Q(m-1/2)(3\cos^2\theta-1+\eta\sin^2\theta\cos2\phi)/2$, where $\nu_Q$, $m$ and $\eta$ respectively denote the quadruple frequency, magnetic quantum number of the initial nuclear energy level and the in-plane EFG anisotropy\cite{Slichter_NMR,Abragam_book}. The $\theta$ and $\phi$ are the pitch and azimuth angle between the applied magnetic field and EFG principle axis.

The single crystals used in this work is of high quality, without any observable disorder effect. The non-zero $\nu_m^{(1)}$ leads to the double satellites, whose line width directly reflects the EFG homogeneity. At $T=240$ K, the full width at half maximum (FWHM) of the satellites is only $\sim24$ kHz for $\mu_0H\perp c$ and $\sim19$ kHz for $\mu_0H||c$, obtained from Lorentz peak fittings. With sample cooling down, the line width broadens gradually. The second order correction to the central line frequency resulting from EFG is less than 3 kHz even for $\mu_0H=10$ T perpendicular to the $c$-axis. The line broadening mainly results from the slight inhomogeneous spin susceptibility at low temperatures. Even for $T=4$ K, the FWHM of the satellites becomes $\sim130$ kHz and $\sim91$ kHz for different field directions. The $\nu_Q$ is calculated to be $\sim1.34$ MHz at $T=40$ K, very close to that in its counterpart Pr$_3$BWO$_9$\cite{Zeng_PRB_104_155150}.

The frequency of the central transition shifts upwards or downwards upon cooling for different field directions. The temperature dependence of the Knight shift is plotted in Fig.\ref{knight2}(c). The Knight shift, defined as relative frequency shift with respect to Larmor frequency, measures the local spin susceptibility. For the 10 T field applied along $c$-axis, the Knight shift decreases from positive values gradually to negative ones with decreasing temperature. Under 10 T in-plane field, the Knight shift first shows an upturn behavior upon cooling, then begins to saturate below $T=2$ K, reflecting the saturated spin susceptibility. This saturation can be seen clearly in the isothermal $M$($H$) curve\cite{Ashtar_IC_59_5368}, which is also widely observed in other frustrated\cite{Zeng_PRB_102_045149,Baenitz_PRB_98_220409,Bordelon_NP_15_1058} or Kitaev antiferromagnets\cite{Zheng_PRL_119_227208}. When decreasing the field intensity to 3.5 T, the upturn behavior restores and persists down to $T=0.4$ K, below which the spin system enters the magnetic ordered state.

The Knight shift comprises a spin part and an orbital part, denoted as $K_{spin}$ and $K_{orb}$, respectively. The spin part $K_{spin}$ can be expressed as $K_{spin}=A_{hf}\chi_s$, where $A_{hf}$ and $\chi_s$ respectively denotes the hyperfine coupling constant and  spin susceptibility. While, the orbital contribution resulting from electron orbital movement is temperature independent. Regarding the enhanced $dc$-susceptibility at low temperatures, the seemingly unusual behavior observed with $\mu_0H||c$ indicates a negative element $A^{cc}_{hf}$ of the hyperfine coupling tensor. In Fig.\ref{knight2} (d), we plot the absolute value of Knight shift versus $dc$-susceptibility with temperature as an explicit parameter(also called Clogston-Jaccarino plot\cite{Clogston_PR_121_1357}). As indicated by the solid straight lines, the linear dependence holds good in a wide range. Thus, the temperature dependence of local spin susceptibility reproduces that of the bulk susceptibility in the paramagnetic state.

\begin{figure}
\includegraphics[width=8cm, height=7.5cm]{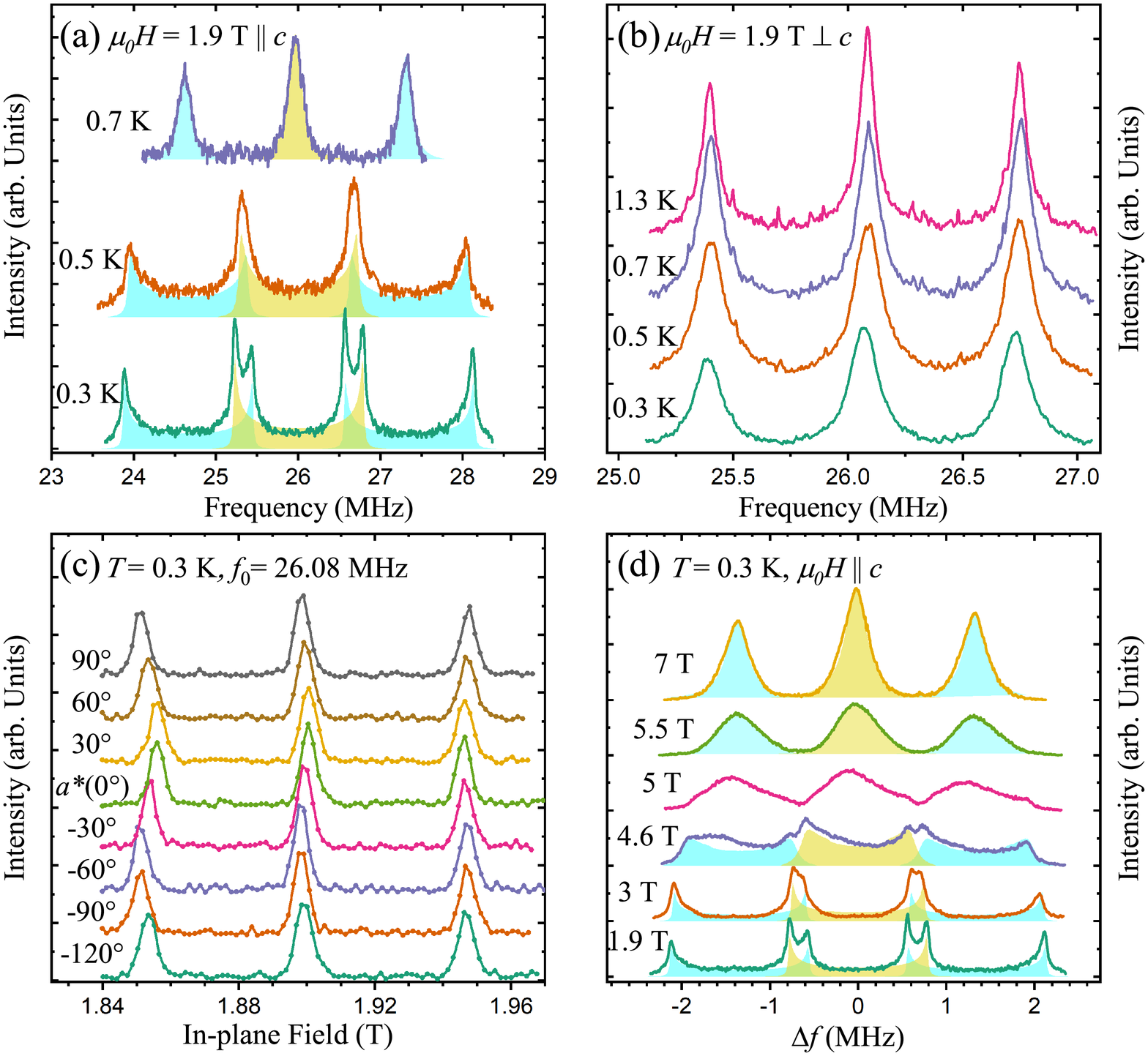}
\caption{\label{afm3}(color online)
  Temperature dependence of the $^{11}$B NMR spectra with a 1.9 T field applied along(a) or perpendicular to $c$-axis(b).
  (c): $^{11}$B NMR spectra at $T=0.3$ K with the field rotated in the $ab$-plane.
  (d): Field dependence of NMR spectra at $T=0.3$ K with $\mu_0H||c$-axis. The $\Delta f$ is defined as the frequency difference from the Larmor frequency. Simulations of the line shape for the incommensurate order are shown as shadowed areas in (a) and (d). The spectral intensity is Boltzmann-corrected to get rid of the influence of temperature and magnetic field.
}
\end{figure}

Next, we turn to study the  magnetic ordered state under low magnetic field. In Fig.\ref{afm3}(a) and (b), we show the $^{11}$B NMR spectra under $\mu_0H=1.9$ T with different directions. For both, all the three peaks are maintained above $T=0.7$ K as the high field case.
Upon cooling, both the central peak and satellite peaks develop into double-horn type peaks under the field along $c$-axis. Surprisingly, the double-horn peaks are absent in the ordered state with the field applied perpendicular to $c$-axis as shown in Fig.\ref{afm3}(b). To eliminate the possibility of coincidence, the spectra at $T=0.3$ K are measured with the field rotated in $ab$-plane(Fig.\ref{afm3}(c)) although the $a$- and $b$-axis are equivalent to each other from the crystal structure. For the present NMR single crystals, the $a$ and $b$-axis are not determined crystallographically. The zero degree is marked with $a^*$, where the central peak gain its frequency maximum among the angle rotations. Expectedly, the double-horn line shape seen under $c$-axis field is absent within the studied field angle range exceeding 180$^o$.

An incommensurate magnetic order with single-$\overrightarrow{q}$ modulation is indicated by our NMR spectra. In NMR, the resonance frequency is determined by total effective field contributed by the vector sum of external applied field and internal hyperfine field, $|\overrightarrow{B}_{total}|=|\overrightarrow{B}_{ext}+\overrightarrow{B}_{int}|$. For the commensurate antiferromagnetic order, each resonance peak splits into double peaks as a result of the doubled magnetic unit cell when $\overrightarrow{B}_{ext}||\overrightarrow{B}_{int}$. If $\overrightarrow{B}_{ext}\perp\overrightarrow{B}_{int}$, the resonance peak will shift to the high frequency side. However, for the incommensurate case with $\overrightarrow{B}_{ext}||\overrightarrow{B}_{int}$, the mismatch between lattice and magnetic period gives birth to the diffusing spectral weight\cite{Ding_PRB_95_184404}. The double-horn type line shape, hallmark of the incommensurate order with single-$\overrightarrow{q}$ modulation\cite{Blinc_PNMRS_41_49}, results from the weighted distribution of the angle between the internal and external field.

We simulate the line shape for the single-$\overrightarrow{q}$ incommensurate modulation as shown by the shadowed area in Fig.\ref{afm3}(a) and (d). The spectral weight contributed by the central transition and satellites is shown separately by different colors. The line shape are simulated by simply adding a cosine-modulated internal field to the applied field, and summing up the weighted spectral intensity together. The intensity scaling between the central line and each satellite is assigned to be $4:3$ according to the theoretical transition probability. The internal field maximum contributed from ordered moments is calculated to be $B_{int}\sim0.1135$ T for $\mu_0H=1.9$ T and $T=0.3$ K. The completely absence of line-splitting under the in-plane field indicates the incommensurate modulated internal field is strictly along the $c$-axis.

With the field intensity increasing, the incommensurate order is gradually suppressed. In Fig.\ref{afm3}(d), we show the full spectra at $T=0.3$ K as a function of $\Delta f$ (the frequency difference from the Larmor frequency) under different field intensities. The double horns become closer to each other with strengthened magnetic field. The internal field maximum is suppressed to $B_{int}\sim0.109$ T at $\mu_0H=3$ T and $B_{int}\sim0.082$ T at $\mu_0H=4.6$ T. The line broadening appears at $\mu_0H=4.6$ T, resulting from the instability of the incommensurate order. When the field strength exceeds $5$ T, the double-horn structure becomes indistinguishable. This indicates the complete suppression of magnetic ordering for the studied temperature range.

\begin{figure}
\includegraphics[width=8cm, height=7.5cm]{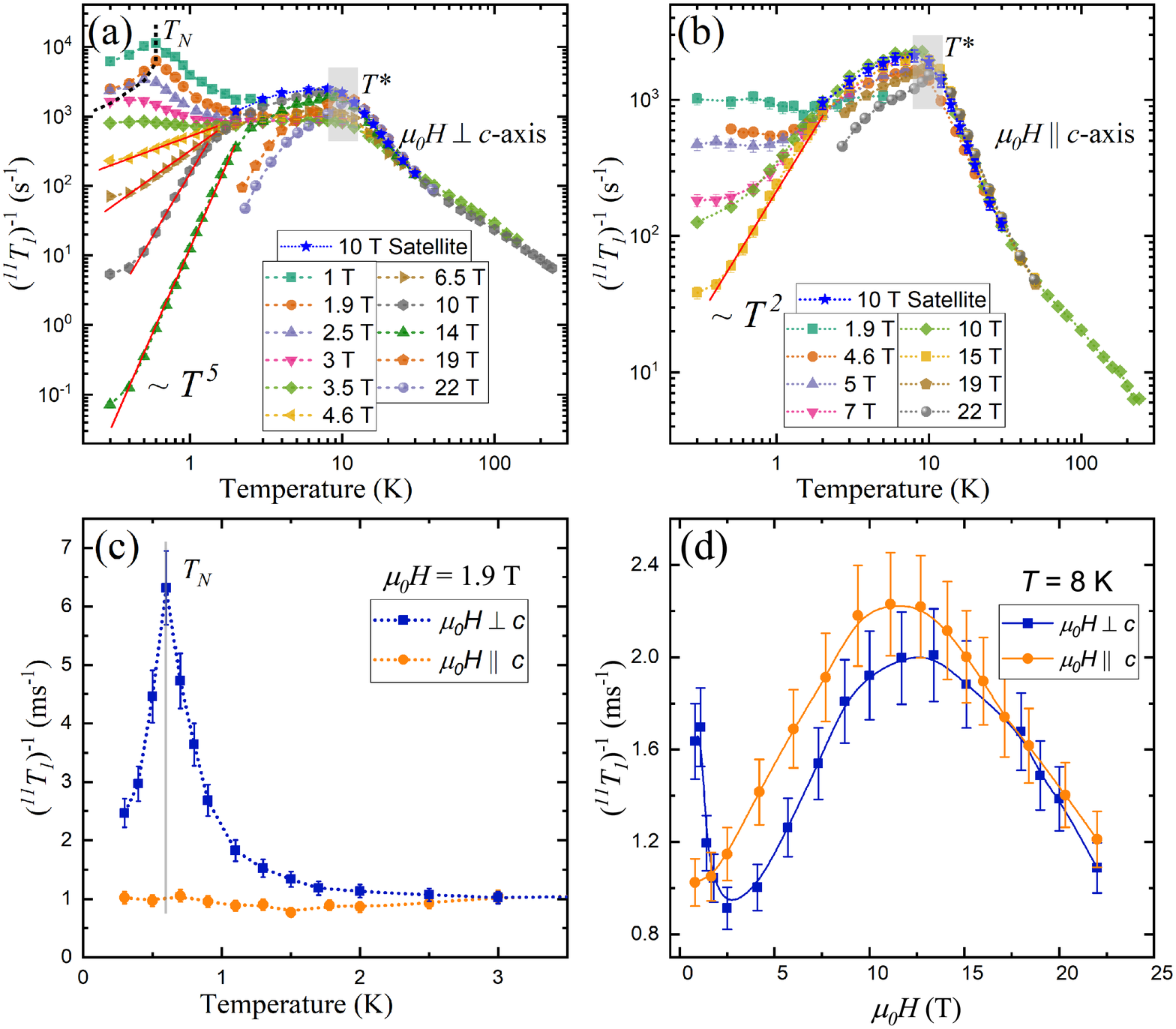}
\caption{\label{SLRR4}(color online)
  The spin-lattice relaxation rate ($1/T_1$) as a function of temperature under different field intensities with $\mu_0H\perp c$(a) or  $\mu_0H||c$-axis(b). The red straight lines show the power-law temperature dependence of $1/T_1$. (c): Temperature dependence of $1/T_1$ under 1.9 T field with different field directions. (d): Field dependence of $1/T_1$ at $T=8$ K under different field directions.
}
\end{figure}

More information about the magnetic order can be obtained from the spin dynamics across the phase transition. We study the low-energy spin dynamics through the nuclear spin-lattice relaxation rate ($1/^{11}T_1$). The stretching behavior is not observed from the recovery curve indicating the highly uniform $1/^{11}T_1$ across the single crystal. The temperature dependence of $1/^{11}T_1$ under $\mu_0H\perp c$-axis is shown in Fig.\ref{SLRR4}(a) for a very wide field intensity range. From near room temperature down to $T\sim10$ K, the $1/^{11}T_1$ steeply increases upon cooling, which is followed closely by a bending over behavior ($2$ K$<T<10$ K). We mark this crossover temperature as $T^*$. The $1/^{11}T_1$ below $T=2$ K shows enormous field dependence. With 1 T field applied, the $1/^{11}T_1$ first rises then drops, forming a sharp peak at $T=0.6$ K. This peak corresponds to the critical slowing down behavior of the spin system across $T_N$. With field intensity increasing, the peak is suppressed to lower temperatures and disappears for $\mu_0H\geq3.5$ T. Then, the $1/^{11}T_1$ shows a power-law temperature dependence. With the field applied along $c$-axis, the $1/^{11}T_1(T)$ behavior (See Fig.\ref{SLRR4}(b))is very similar with that under $\mu_0H\perp c$-axis for $2$ K$<T<10$ K. However, what is quite different is that the critical slow down behavior near $T_N$ is absent under this field direction. In Fig.\ref{SLRR4}(c), we specially plot $1/^{11}T_1(T)$ under 1.9 T with different directions in the linear scale, where the contrasting behavior is very obvious.

The spin-lattice relaxation measures the hyperfine field fluctuating intensity in the plane perpendicular to the magnetic field\cite{Slichter_NMR,Abragam_book,Moriya_JPSJ_18_516}. The absence of the critical slowing down behavior with $\mu_0H||c$-axis indicates that the hyperfine field fluctuation is also strictly along the $c$-axis, and the in-plane fluctuation is absent. With increasing field strength, the spin degeneracy is further released, leading to the suppressed  $1/^{11}T_1$ at low temperatures. The power-law behavior should contributes from spin wave like excitations with the gap size smaller than 0.3 K.

The field dependence of $1/^{11}T_1$ at $T=8$ K is further checked, and shown in Fig.\ref{SLRR4}(d). For $\mu_0H||c$-axis, the $1/^{11}T_1$ first rises, then drops with the increasing field intensity. The in-plane field dependence of $1/^{11}T_1$ shares a similar behavior, except the steep rise with the field going down below 2.5 T, reflecting the suppression of spin fluctuations under magnetic field. The temperature and field intensity($\mu_0H>2.5$ T) dependence of $1/^{11}T_1$ around $T^*$ show an nearly isotropic behavior with the field direction. This indicates that the crossover at $T^*$ should have little connection with the incommensurate order. In Fig.\ref{SLRR4}(a) and (b), we also show the $1/^{11}T_1$ measured at the satellite for temperatures around $T^*$, which is more sensitive to the crystal electric field fluctuations. However, its temperature dependence highly coincide with that measured at the central peak, indicating that the hump behavior should also have a magnetic origin.

In rare-earth based frustrated magnets, the magnetic properties arise from the $4f$-electrons in the inner shell of rare-earth ions. The local crystal field at RE$^{3+}$ sites determines the splitting of $4f$ energy levels, also the effective $J$. The upturn behavior of $1/^{11}T_1$ with the sample cooling from room temperature to $T^*$ results from the slowing down of spin fluctuations contributed by the thermally activated Sm$^{3+}$ multiplet with stronger correlations. Below $T^*$, the thermally excited state is frozen out, and the ground state of $4f$ energy levels with $J_{eff}$ dominate the low temperature magnetic properties of the present sample. This behavior is also observed in its counterpart Pr$_3$BWO$_9$\cite{Zeng_PRB_104_155150} and other frustrated rare-earth materials \cite{Xu_PRL_103_267402,Li_PRL_118_107202,Zeng_PRB_102_045149}.
The transitions between multiplets of localized $f$-electrons may count for the field dependent spin fluctuations above 2.5 T. More information about the specific crystal field splitting needs further high-energy neutron scattering measurements.

\begin{figure}
\includegraphics[width=7cm, height=7cm]{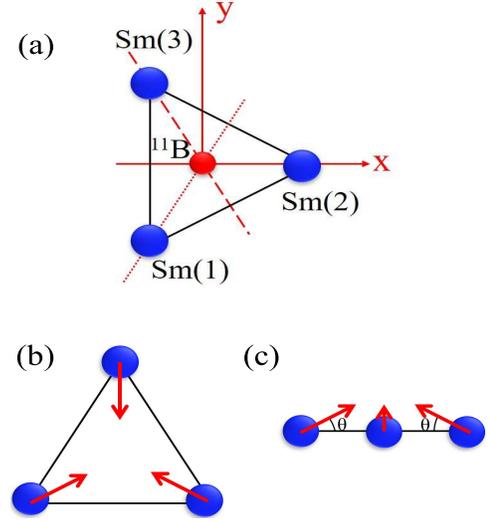}
\caption{\label{HCT}(color online)
  (a) A sketch of the studied $^{11}$B nuclear sites and three nearest neighboring Sm$^{3+}$ as seen against the $c$-axis. Planes containing $z$($c$)-axis and $x$-axis, dotted line or dashed line are three mirror planes.
  (b) and (c): Constraints on the possible magnetic structure as seen perpendicular to or parallel with the kagome plane (See the text).
}
\end{figure}

Next, we discuss possible magnetic structure of the incommensurate order in Sm$_3$BWO$_9$. As the magnetic moments contribute to the internal field via hyperfine coupling, the coupling tensor is very important for the analysis. In Fig.\ref{HCT}(a), we sketch the three nearest Sm$^{3+}$ neighbors of the $^{11}$B site. The $^{11}$B sits right above the center of the regular triangle formed by three Sm$^{3+}$ magnetic ions, which are numbered as Sm(1) to Sm(3). Obviously, the symmetry operations about the studied $^{11}$B nuclear site are exactly the C$_\text{3v}$ group. The hyperfine coupling tensors can be generally written as,
$$
\left(
  \begin{array}{ccc}
      A_{11}&A_{12}&A_{13}\\
      A_{21}&A_{22}&A_{23}\\
      A_{31}&A_{32}&A_{33}\\
  \end{array}
\right)
\left(
  \begin{array}{ccc}
      B_{11}&B_{12}&B_{13}\\
      B_{21}&B_{22}&B_{23}\\
      B_{31}&B_{32}&B_{33}\\
  \end{array}
\right)
\left(
  \begin{array}{ccc}
      C_{11}&C_{12}&C_{13}\\
      C_{21}&C_{22}&C_{23}\\
      C_{31}&C_{32}&C_{33}\\
  \end{array}
\right),
$$
respectively for $^{11}$B-Sm(1), $^{11}$B-Sm(1) and $^{11}$B-Sm(3).
The hyperfine field $\mu_0\overrightarrow{H^1}$contributed by the magnetic moment at Sm(1) $\overrightarrow{M^1}$ can be expressed as,
\begin{equation}
\mu_0\left(
  \begin{array}{c}
      H^1_x\\
      H^1_y\\
      H^1_z\\
  \end{array}
\right)
=
\left(
  \begin{array}{ccc}
      A_{11}&A_{12}&A_{13}\\
      A_{21}&A_{22}&A_{23}\\
      A_{31}&A_{32}&A_{33}\\
  \end{array}
\right)
\left(
  \begin{array}{c}
      M^1_x\\
      M^1_y\\
      M^1_z\\
  \end{array}
\right).
\end{equation}
For Sm(2) and Sm(3), we can also have,
\begin{equation}
\mu_0\left(
  \begin{array}{c}
      H^2_x\\
      H^2_y\\
      H^2_z\\
  \end{array}
\right)
=
\left(
  \begin{array}{ccc}
      B_{11}&B_{12}&B_{13}\\
      B_{21}&B_{22}&B_{23}\\
      B_{31}&B_{32}&B_{33}\\
  \end{array}
\right)
\left(
  \begin{array}{c}
      M^2_x\\
      M^2_y\\
      M^2_z\\
  \end{array}
\right),
\end{equation}
and
\begin{equation}
\mu_0\left(
  \begin{array}{c}
      H^3_x\\
      H^3_y\\
      H^3_z\\
  \end{array}
\right)
=
\left(
  \begin{array}{ccc}
      C_{11}&C_{12}&C_{13}\\
      C_{21}&C_{22}&C_{23}\\
      C_{31}&C_{32}&C_{33}\\
  \end{array}
\right)
\left(
  \begin{array}{c}
      M^3_x\\
      M^3_y\\
      M^3_z\\
  \end{array}
\right).
\end{equation}

We can reduce the hyperfine coupling tensor based on the symmetry operations. From Sm(1) to Sm(2), two symmetry operations can be performed: one is the mirror symmetry about the plane containing the dashed line and $z$-axis in Fig.\ref{HCT}; the other one is counterclockwise rotating by 120$^o$ about $z$-axis. By the former operation, we can get the equation,
\begin{equation}
\begin{split}
&\mu_0\left(
  \begin{array}{ccc}
      -\frac{1}{2}&-\frac{\sqrt{3}}{2}&0\\
      -\frac{\sqrt{3}}{2}&\frac{1}{2}&0\\
      0&0&1
  \end{array}
\right)
\left(
  \begin{array}{c}
      H^1_x\\
      H^1_y\\
      H^1_z\\
  \end{array}
\right)\\
&=
\left(
  \begin{array}{ccc}
      B_{11}&B_{12}&B_{13}\\
      B_{21}&B_{22}&B_{23}\\
      B_{31}&B_{32}&B_{33}\\
  \end{array}
\right)
\left(
  \begin{array}{ccc}
      -\frac{1}{2}&-\frac{\sqrt{3}}{2}&0\\
      -\frac{\sqrt{3}}{2}&\frac{1}{2}&0\\
      0&0&1
  \end{array}
\right)
\left(
  \begin{array}{c}
      M^1_x\\
      M^1_y\\
      M^1_z\\
  \end{array}
\right).
\end{split}
\end{equation}
By the second operation, we can obtain,
\begin{equation}
\begin{split}
&\mu_0\left(
  \begin{array}{ccc}
      -\frac{1}{2}&-\frac{\sqrt{3}}{2}&0\\
      \frac{\sqrt{3}}{2}&-\frac{1}{2}&0\\
      0&0&1
  \end{array}
\right)
\left(
  \begin{array}{c}
      H^1_x\\
      H^1_y\\
      H^1_z\\
  \end{array}
\right)\\
&=
\left(
  \begin{array}{ccc}
      B_{11}&B_{12}&B_{13}\\
      B_{21}&B_{22}&B_{23}\\
      B_{31}&B_{32}&B_{33}\\
  \end{array}
\right)
\left(
  \begin{array}{ccc}
      -\frac{1}{2}&-\frac{\sqrt{3}}{2}&0\\
      \frac{\sqrt{3}}{2}&-\frac{1}{2}&0\\
      0&0&1
  \end{array}
\right)
\left(
  \begin{array}{c}
      M^1_x\\
      M^1_y\\
      M^1_z\\
  \end{array}
\right).
\end{split}
\end{equation}
As the tensor $\{B_{ij}\}$ simultaneously satisfies Equation (4) and (5) for any given magnetic moment $\overrightarrow{M^1}$, the matrix elements $B_{12}$, $B_{21}$, $B_{23}$ and $B_{32}$ are forced to be zero. By performing other symmetry operations in the $c_{3v}$-group, the hyperfine coupling tensors are calculated to be as follows,\\
For $^{11}$B-Sm(1): $\{A_{ij}\}=$
\begin{equation}
\left(
  \begin{array}{ccc}
      B_{11}&0&-\frac{1}{2}B_{13}\\
      0&B_{11}&-\frac{\sqrt{3}}{2}B_{13}\\
      -\frac{1}{2}B_{31}&-\frac{\sqrt{3}}{2}B_{31}&B_{33}\\
  \end{array}
\right);
\end{equation}
For $^{11}$B-Sm(2): $\{B_{ij}\}=$
\begin{equation}
\left(
  \begin{array}{ccc}
      B_{11}&0&B_{13}\\
      0&B_{11}&0\\
      B_{31}&0&B_{33}\\
  \end{array}
\right);
\end{equation}
For $^{11}$B-Sm(3): $\{C_{ij}\}=$
\begin{equation}
\left(
  \begin{array}{ccc}
      B_{11}&0&-\frac{1}{2}B_{13}\\
      0&B_{11}&\frac{\sqrt{3}}{2}B_{13}\\
      -\frac{1}{2}B_{31}&\frac{\sqrt{3}}{2}B_{31}&B_{33}\\
  \end{array}
\right).
\end{equation}

By substituting the general hyperfine coupling tensors in Equation (1)-(3) with the reduced ones shown in (6)-(8), the in-plane component of the internal field can be obtained,
\begin{equation}
\begin{split}
\mu_0H_x=&B_{11}(M^1_x+M^2_x+M^3_x)\\
&-\frac{1}{2}B_{13}M^1_z+B_{13}M^2_z-\frac{1}{2}B_{13}M^3_z,
\end{split}
\end{equation}

\begin{equation}
\begin{split}
\mu_0H_y=&B_{11}(M^1_y+M^2_y+M^3_y)\\
&-\frac{\sqrt{3}}{2}B_{13}M^1_z+\frac{\sqrt{3}}{2}B_{13}M^3_z.
\end{split}
\end{equation}

Our NMR results place two constraints to the possible magnetic structure. In the magnetically ordered state of Sm$_3$BWO$_9$, the internal hyperfine field and its fluctuation are strictly along the crystalline $c$-axis. To satisfy this observation, the in-plane components of the magnetic moments on the regular triangle must make a 120$^o$ angle (Fig.\ref{HCT}(b)), and the $c$-axis components, if are not zero, must equal with each other(Fig.\ref{HCT}(c)). If the magnetic moments lies on the $ab$-plane, the incommensurability can only be realized via modulating the magnetic moment size, yielding the spin density wave state. Otherwise, the incommensurate order can be realized by continuously tuning the angle between the magnetic moment and the $c$-axis, modulating the magnetic moment size, or both of them. Further neutron scattering study is needed to finally determine the magnetic structure.

We close our discussions by comparing the ground state of Sm$_3$BWO$_9$ with that of Pr$_3$BWO$_9$ from the same family. In Pr$_3$BWO$_9$, the fluctuating paramagnetic state persist down to $T=0.09$ K without any magnetic ordering or spin freezing\cite{Zeng_PRB_104_155150}. From the spin dynamics, a moderate spin excitation gap is observed by NMR, whose gap size is proportional to the applied magnetic field intensity. Comparatively, the spin system in Sm$_3$BWO$_9$ enters magnetically ordered state below $T_N=0.75$ K under zero magnetic field. The absence of thermally activation behavior in $1/T_1$($T$) at low temperatures indicates the magnon excitation gap, if exist, should be smaller than 0.3 K. The single ion anisotropy should play an important role in the determination of such contrasting ground states. The uniaxial anisotropy of the Pr$^{3+}$-spin is indicated by the negative $D$ in Pr$_3$BWO$_9$, and the $S^z=\pm1$ state is energetically favorable. While, the spin anisotropy of the Sm$^{3+}$-spin is very weak as evidenced by the gapless spin excitations. The magnetically ordered state in Sm$_3$BWO$_9$ should be selected out of the highly degenerated states by the thermal fluctuations via the order-by-disorder mechanism.

To conclude, we employ NMR as a local probe to investigate the magnetic ground state of Sm$_3$BWO$_9$ with distorted kagome lattice. An incommensurate magnetic order is shown by the double-horn type spectra under a $c$-axis magnetic field. While, absence of line splitting is observed with an in-plane field, indicating the incommensurate modulation of the internal field strictly along $c$-axis. From the spin dynamics, the critical slowing down behavior is observed in the temperature dependence of $1/T_1$ with $\mu_0H\perp c$-axis, which is completely absent in that with $\mu_0H || c$-axis.
Based on the local symmetry of $^{11}$B sites, we analyze the hyperfine coupling tensors, and further propose two constraints on the possible magnetic structure. The contrasting magnetic ground states of Pr$_3$BWO$_9$ and Sm$_3$BWO$_9$ should be related with the diverse single ion anisotropy.

This research was supported by the National Natural Science Foundation of China (Grants No. 11874057, 11874158, U1732273 and 21927814). A portion of this work was supported by the High Magnetic Field Laboratory of Anhui Province.


\end{document}